\documentclass{article}
\usepackage{romp}

\usepackage[dvips]{graphicx}
\usepackage{amsmath}
\usepackage{amsfonts}
\usepackage{amssymb}
\usepackage{longtable} 
\usepackage{color}

\allowdisplaybreaks[4] 

\newcommand{\Chr}[3]{\mbox{\small\( \begin{Bmatrix}#1\\#2#3\end{Bmatrix}\)}}
\newcommand{\tChr}[3]{\mbox{\scriptsize\(\begin{Bmatrix}#1\\#2#3\end{Bmatrix}\)}}
\newcommand{\df}{\ {\overset {\rm def} =}\ }
\newcommand{\dr}[2]{\frac {{\rm d} {#1}} {{\rm d} {#2}}}
\newcommand{\pdr}[2]{\frac {\partial {#1}} {\partial {#2}}}
\newcommand{\dril}[2]{{{\rm d} {#1}} / {{\rm d} {#2}}}

\title{Causality in the maximally extended extreme Reissner--Nordstr\"{o}m
spacetime with identifications}
\author{Andrzej Krasi\'nski \\ N. Copernicus Astronomical Centre, Polish Academy
of Sciences \\
Bartycka 18, 00 716 Warszawa, Poland \\
e-mail: akr@camk.edu.pl}

\begin{document}

\maketitle
\begin{abstract}
In continuation of the similarly titled paper on the $e^2 < m^2$ Reissner --
Nordstr\"{o}m (RN) metric, in this paper it was verified whether it is possible
to send (by means of timelike and null geodesics) messages to one's own past in
the maximally extended {\it extreme} ($e^2 = m^2$) RN spacetime with the
asymptotically flat regions being identified. Numerical examples show that
timelike and nonradial null geodesics originating outside the horizon have their
turning points to the future of the past light cone of the future copy of the
emitter. This means that they cannot reach the causal past of the emitter's
future copy. Ingoing radial null geodesics hit the singularity at $r = 0$ and
stop there. So, unlike in the $e^2 < m^2$ case, identification of the
asymptotically flat regions does not lead to causality breaches. A formal
mathematical proof of this thesis (as opposed to the numerical examples given in
this paper) is still lacking and desired.
\end{abstract}

\section{Motivation and summary}\label{intro}

This paper is a continuation of Ref. \cite{Kras2025}. In that one it was shown
that in the maximally extended Reissner \cite{Reis1916} -- Nordstr\"{o}m
\cite{Nord1918} (RN) spacetime with $e^2 < m^2$ and with the consecutive
asymptotically flat regions being identified, it is possible to send messages to
one's own past by means of radial timelike geodesics, provided the message is
emitted early enough. Thus, the identifications lead to causality breaches.

In the present paper the same problem was investigated for the extreme ($e^2 =
m^2$) RN spacetime with the asymptotically flat regions identified. Numerical
integrations of the geodesic equations showed that for timelike and nonradial
null geodesics their turning points lie to the future of the past light cone of
the copy of the initial point. This means that the geodesics can come back to
the copy of the observer's worldline only later than they were emitted, i.e. no
breach of causality is caused by the identifications. Now it remains an open
problem to prove this thesis by formal mathematical arguments (as opposed to
numerical examples given here).

In Sec. \ref{RNintro} the basic facts about the extreme RN metric are recalled,
and its maximal extension is re-derived, including a few details that are
omitted in textbook presentations. In suitably chosen coordinates, the image of
the singularity at $r = 0$ is a straight line \cite{Cart1966}.

In Sec. \ref{geodesics}, the equations of timelike and null geodesics in the
extreme RN metric are discussed. It is shown that ingoing radial null geodesics
hit the singularity at $r = 0$ and stop there. For a radial timelike geodesic,
the $r$ coordinate of its turning point is explicitly calculated. It is also
shown that timelike and nonradial null geodesics are tangent to the horizon in
the $(U, V)$ coordinates of the maximal extension.

In Sec. \ref{numgeo}, numerical examples of radial timelike geodesics are
investigated. They show that the turning points of these geodesics lie to the
future of the past light cone of the first future copy of the initial point.
This means that such geodesics cannot carry messages to the past of their
emitters and so do not break causality.

In Sec. \ref{norad}, nonradial timelike and null geodesics originating in an
asymptotically flat region were investigated and the positions of their turning
points (TPs) were discussed. For null geodesics, these positions were explicitly
calculated. One TP, inside the horizon, exists for all values of the energy
($\Gamma$) and nonzero angular momentum ($J_0 \neq 0$) constants. For
sufficiently large $|J_0|$, two (in a limiting case one) extra TPs exist outside
the horizon. In this case a light ray can either propagate between the outermost
TP and infinity (these are irrelevant for the problem of causality) or oscillate
between the other two TPs, passing all through the black hole in each cycle.

In Sec. \ref{numnorad}, examples of nonradial timelike and null geodesics that
cross the horizon are numerically integrated. Together with the examples from
Sec. \ref{numgeo} they show that the TPs of timelike and null geodesics lie to
the future of the past light cones of the future copies of their initial points,
and so no breach of causality occurs. The null geodesics that have large $|J_0|$
and go off a point between the horizon and the middle TP should go through the
black hole to the next asymptotically flat region. However, the numerical
integration of their paths beyond the horizon was impossible in consequence of
the extreme compression of the large-$r$ regions involved in the transformation
to the coordinates of the maximal extension.

In Sec. \ref{conclu}, conclusions are presented.

Six appendices present details of selected reasonings and calculations.

\section{Basic facts about the maximally extended extreme RN
metric}\label{RNintro}

\setcounter{equation}{0}

The signature $(+ - - -)$ will be used throughout the paper.

The extreme RN metric is a spherically symmetric electrovacuum solution of the
Einstein -- Maxwell equations that describes the vicinity of a body (or black
hole) of mass $m$ and electric charge $e$ such that $|e| = m$. It is called
extreme because it has the largest $|e| / m$ ratio, for which a horizon still
exists (with $|e| < m$ there are two horizons that merge into one when $|e| \to
m$, and there is no horizon when $|e| > m$). In curvature
coordinates\footnote{See Ref. \cite{PlKr2024} for the RN metric expressed in the
Lema\^{\i}tre \cite{Lema1933} -- Novikov \cite{Novi1964,KEMa2013} coordinates.}
it is
\begin{equation}\label{2.1}
{\rm d} s^2 = \phi\ {\rm d} t^2 - {\rm d} r^2 / \phi - r^2 \left({\rm d}
\vartheta^2 + \sin^2 \vartheta\ {\rm d} \varphi^2\right), \qquad \phi \df  (1 -
m / r)^2.
\end{equation}
The mass $m$ and the charge $e$ are expressed in units of length. They are
related to the mass $M$ and charge $Q$ in physical units by $m = GM/c^2$ and $e
= \sqrt{G} Q / c^2$, where $G$ is the gravitational constant and $c$ is the
velocity of light (see Eq. (19.62) in Ref. \cite{PlKr2024}). The metric
(\ref{2.1}) has a spurious singularity at $r = m$, see Appendix \ref{spusing}.
The singularity at $r = 0$ is genuine because there the scalar components of the
Riemann tensor diverge.

\begin{figure}[h]
 \begin{center}
 \includegraphics[scale=0.8]{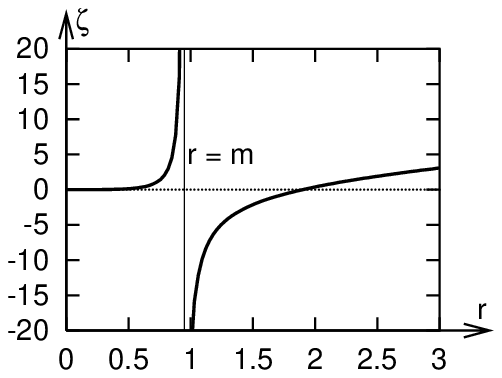}
 \caption{
 \label{tworanges}
 \footnotesize
The graph of the function $\zeta(r)$ defined by (\ref{2.2}), with $m = 0.95$
(the same value as in Ref. \cite{Kras2025}).}
 \end{center}
 \end{figure}

We transform the $r$ coordinate by
\begin{equation}\label{2.2}
\zeta \df \int \frac {{\rm d} r} {\phi} = r - m - \frac {m^2} {r - m} + 2m
\ln|r/m - 1|,
\end{equation}
see Appendix \ref{major} for remarks on the inverse function $r(\zeta)$. The
transformed metric is
\begin{equation}\label{2.3}
{\rm d} s^2 = \left[(1 - m / r)^2 \left({\rm d} t^2 - {\rm d} \zeta^2\right) -
r^2 \left({\rm d} \vartheta^2 + \sin^2 \vartheta {\rm d}
\varphi^2\right)\right]_{r = r(\zeta)}.
\end{equation}
In these coordinates there is no singularity in the Christoffel symbols at $r =
m$, see Appendix \ref{Christ} (so no singularity in the geodesic equations). We
have (see Fig. \ref{tworanges}):
\begin{equation}\label{2.4}
\zeta(0) = 0, \quad \lim_{r \to m^-} \zeta(r) = + \infty, \quad \lim_{r \to m^+}
\zeta(r) = - \infty, \quad \lim_{r \to +\infty} \zeta(r) = + \infty.
\end{equation}
The domain $r \in (0, +\infty)$ is thus covered by two $(t, \zeta)$ coordinate
patches, one for $r \in (0, m)$ and the other for $r \in (m, +\infty)$. The
inverse function $r(\zeta)$ is uniquely defined in each of the ranges $r \in (0,
m)$ and $r \in (m, +\infty)$ since $\dril \zeta r > 0$ for all $r$.

We now transform $(t, \zeta)$ to the null coordinates
\begin{equation}\label{2.5}
p = t - \zeta, \qquad q = t + \zeta,
\end{equation}
and then to\footnote{The $(P, Q)$ in (\ref{2.6}) are different from the $(P, Q)$
of (14.170) in Ref. \cite{PlKr2024}.}
\begin{equation}\label{2.6}
(P, Q) = (\tanh p, \tanh q) \Longleftrightarrow (p, q) = \left(\frac 1 2\ \ln
\frac {1 + P} {1 - P}, \frac 1 2\ \ln \frac {1 + Q} {1 - Q}\right).
\end{equation}
The $(P, Q)$ have the ranges $P \in [-1, +1]$, $Q \in [-1, +1]$, with $r \to +
\infty$ corresponding to $P = -1$ (past null infinity) and $Q = +1$ (future null
infinity). The set $r = m$ has the equation $\{P = -1\} \cup \{Q = +1\}$ in the
$r \leq m$ patch and $\{P = +1\} \cup \{Q = -1\}$ in the $r \geq m$ patch. At
the singularity $r = 0$ we have $\zeta = 0$, so $P = Q$.

It is convenient to introduce the space -- time coordinates $(U, V)$ by
\begin{equation}\label{2.7}
U = (Q - P) / 2, \qquad V = (P + Q) / 2,
\end{equation}
of which $V$ is timelike and $U$ is spacelike. From the properties  of $P$ and
$Q$ we see that
\begin{equation}\label{2.8}
|U| \leq 1, \quad |V| \leq 1, \qquad |U + V| \leq 1, \quad |U - V| \leq 1.
\end{equation}
The singularity now lies on the $V$ coordinate axis. The set $r = m$ belongs to
the $r =$ constant family, that is to
\begin{equation} \label{2.9}
p - q = - 2 \zeta \df C = {\rm constant}
\end{equation}
and corresponds to $C \to - \infty$ on the $r < m$ side and $C \to + \infty$ on
the $r > m$ side. By (\ref{2.6}) and (\ref{2.7}), $p - q = C$ is equivalent to
\begin{equation}\label{2.10}
\frac {(1 - U)^2 - V^2} {(1 + U)^2 - V^2} = {\rm e}^{2C} \Longleftrightarrow V^2
= U^2 + 2\ \frac {\cosh C} {\sinh C}\ U + 1.
\end{equation}
The lines of constant $r$ are thus hyperbolae in the $(U, V)$ plane with
vertices on the $U$ axis. From the above, we obtain for $r \to m^-$, i.e. for $C
\to - \infty$
\begin{equation}\label{2.11}
V^2 = (1 - U)^2 \Longrightarrow V = 1 - U \quad {\rm and} \quad V = U - 1,
\end{equation}
and for $r \to m^+$, i.e. for $C \to + \infty$
\begin{equation}\label{2.12}
V^2 = (1 + U)^2 \Longrightarrow V = 1 + U \quad {\rm and} \quad V = - U - 1.
\end{equation}

 \begin{figure}[h]
 ${}$ \\[-3.5cm]
 \includegraphics[scale=0.7]{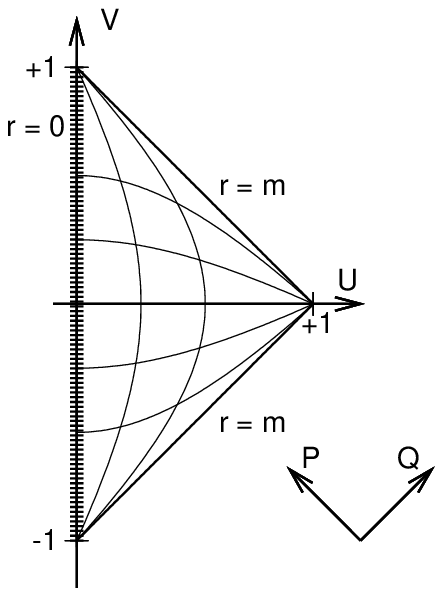}
 \hspace{-1cm} \includegraphics[scale=0.7]{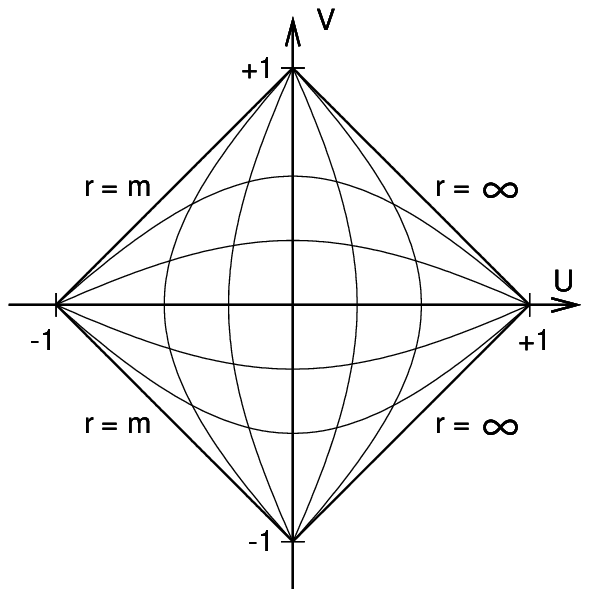}
 \caption{
 \label{compact}
 \footnotesize
The subsets of the $(U, V)$ coordinate plane corresponding to $r \leq m$ (left
panel) and $r \geq m$ (right panel). The vertical hyperbolae are the $r =$
constant lines, they are timelike. They degenerate to the pairs of straight
segments (which are null) in the limits $r \to m$ and $r \to \infty$. The
horizontal hyperbolae are the $t =$ constant lines. }
 \end{figure}

\begin{figure}
 \begin{center}
 ${}$ \\[-5mm]
 \includegraphics[scale=0.8]{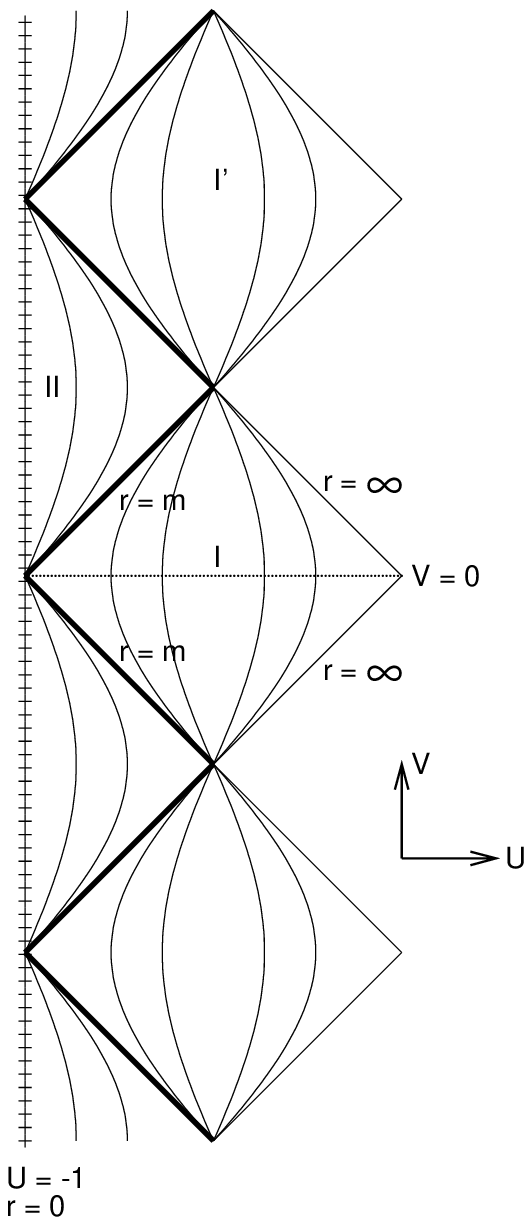}
 ${}$ \\[5mm]
 \caption{
 \label{rnextmax}
 \footnotesize
The maximal extension of the extreme ($e^2 = m^2$) R--N metric. The thin
straight segments are the images of the null infinities, where $r \to \infty$.
The hyperbola arcs are the timelike $r =$ constant $\neq m$ lines. The thick
straight segments are the spurious singularities (event horizons) at $r = m$.
The hatched straight line is the true singularity at $r = 0$; it coincides with
the ${\cal U} = -1$ coordinate line and is timelike. Just as in the $e^2 < m^2$
case, we can identify sectors I and I$'$. }
 \end{center}
 \end{figure}

The subset of the $(U, V)$ plane delimited by the lines (\ref{2.11}) includes $r
= 0$, i.e. $U = 0$. The subset delimited by (\ref{2.12}) includes $p - q \to -
\infty$ at $r \to + \infty$. These two subsets are shown in Fig. \ref{compact}.
In constructing the maximal extension we lay them side by side so that the $r =
m$ lines coincide. The result is shown in Fig. \ref{rnextmax}. The image of the
$r = 0$ singularity in the $({\cal U}, {\cal V})$ plane is the straight line
${\cal U} = -1$, unlike in the $e^2 < m^2$ case \cite{Kras2025}, where the shape
of this line depended on $m$ and $e$. A diagram equivalent to Fig.
\ref{rnextmax} was first presented by Carter \cite{Cart1966,Cart1973}. The
$({\cal U}, {\cal V})$ coordinates in Fig. \ref{rnextmax} coincide with the
internal $(U, V)$ coordinates of (\ref{2.7}) only in sector I. In other sectors,
$({\cal U}, {\cal V})$ are shifted with respect to $(U, V)$, for example in
sector II $({\cal U}, {\cal V}) = (U - 1, V + 1)$. Similarly to the $e^2 < m^2$
case, we can identify sector I$'$ with sector I. This could possibly lead to
breaches of causality, and this possibility is the main subject of this paper.

By the same method as above we conclude that in the $(U, V)$ coordinates the
lines of constant $t = (p + q) / 2 \df D$ are hyperbolae with the vertices on
the $V$ axis given by
\begin{equation}\label{2.13}
U^2 = V^2 - 2 \frac {\cosh D} {\sinh D}\ V + 1.
\end{equation}
In the limit $D \to + \infty$ they coincide with the upper $r = m$ line in the
$r \leq m$ sector and with the upper $r = m$ and $r = \infty$ lines in the $r
\geq m$ sector. In the limit $D \to - \infty$, they coincide with the lower $r =
m$ line in the $r \leq m$ sector and with the lower $r = m$ and $r = \infty$
lines in the $r \geq m$ sector.

\section{The geodesics in the metric (\ref{2.1})}\label{geodesics}

\setcounter{equation}{0}

Coordinates may be adapted to each single geodesic so that it lies in the
$\vartheta = \pi/2$ hypersurface of (\ref{2.1}) \cite{Kras2025,PlKr2024}. Then
the geodesic equations in the metric (\ref{2.1}) have the following first
integrals \cite{Kras2025}:
\begin{eqnarray}
\phi \dril t s &=& \Gamma, \label{3.1} \\
\left(\dril r s\right)^2 &=& \Gamma^2 - E \phi, \label{3.2} \\
E &\df& \varepsilon + {J_0}^2 / r^2, \label{3.3} \\
\dril {\varphi} s &=& J_0 / r^2, \label{3.4}
\end{eqnarray}
where $\Gamma$ and $J_0$ are arbitrary constants, $\varepsilon = +1$ for
timelike and $\varepsilon = 0$ for null geodesics (spacelike geodesics, on which
$\varepsilon = -1$, will not be considered). With $J_0 = 0$ the geodesic is
radial, with $\Gamma > 0$ ($\Gamma < 0$) it is future- (past-) directed (with
$\Gamma = 0$ it must be spacelike). By virtue of (\ref{3.2}) a timelike geodesic
can reach $r \to \infty$ only when $\Gamma^2 \geq 1$. Null geodesics can reach
$r \to \infty$ with any $\Gamma$ (see, however, Sec. \ref{norad}: whether they
actually reach infinity depends on $J_0$ and the initial point). Timelike and
nonradial null geodesics can run only where $\Gamma^2 - E \phi > 0$ and have
turning points where $\Gamma^2 - E \phi = 0$. For a radial timelike geodesic,
the solution of $\Gamma^2 - E \phi = 0$ is
\begin{equation}\label{3.5}
r_{{\rm TP}\pm} = \frac m {1 \pm |\Gamma|}.
\end{equation}
This shows that hyperbolic or parabolic\footnote{In analogy to Newtonian orbits,
we call a timelike geodesic `hyperbolic' when its equation allows the coordinate
$r$ to go to infinity with $\lim_{r \to \infty} |\dril r s| > 0$, and `elliptic'
when $r$ is bounded from above.} ($|\Gamma| \geq 1$) radial timelike geodesics
have only one turning point (at $r = r_{{\rm TP}+}$), which is inside the
horizon (at $r < m / 2$ when $|\Gamma| > 1$ and at $r = m / 2$ when $|\Gamma| =
1$). The elliptic ($0 < |\Gamma| < 1$) geodesics have one turning point inside
(at $r = r_{{\rm TP}+}$) and the other outside the horizon.

Using (\ref{2.2}), Eq. (\ref{3.2}) is equivalent to
\begin{equation}\label{3.6}
\phi\ \dril {\zeta} s = \sigma \sqrt{\Gamma^2 - E \phi},
\end{equation}
where $\sigma = +1$ for outgoing and $\sigma = -1$ for ingoing geodesics. Thus,
on a radial null geodesic (for which $E = 0$),
\begin{equation}\label{3.7}
\left.\phi\ \dril {\zeta} s\right|_{rn} = \pm \Gamma.
\end{equation}
{}From here and (\ref{3.1}), the equation of a radial null geodesic in the $(t,
\zeta)$ coordinates is
\begin{equation}\label{3.8}
\dril t {\zeta} = \pm 1 \Longrightarrow t \pm \zeta = {\rm constant}.
\end{equation}
Via (\ref{2.5}) -- (\ref{2.7}), Eq. (\ref{3.8}) shows that radial null geodesics
obey $U \pm V =$ constant, i.e. in Figs. \ref{compact} and \ref{rnextmax} they
are straight lines parallel to the $r = m$ lines. They can be extended to
arbitrary values of $\zeta$, so they can hit the singularity at $r = 0$ (and
stop there).

{}From (\ref{2.6}), (\ref{2.5}) and (\ref{2.2}) we find for $\dril P s$ and
$\dril Q s$ along a geodesic:
\begin{eqnarray}
\dr P s &=& \pdr P t\ \dr t s + \pdr P r\ \dr r s = \frac {1 - P^2} {\phi}\
\left(\Gamma - \sigma {\sqrt{\Gamma^2 - E \phi}}\right), \label{3.9} \\
\dr Q s &=& \pdr Q t\ \dr t s + \pdr Q r\ \dr r s = \frac {1 - Q^2} {\phi}\
\left(\Gamma  + \sigma {\sqrt{\Gamma^2 - E \phi}}\right). \label{3.10}
\end{eqnarray}
Suppose we choose an initial point E1 in sector I of Fig. \ref{rnextmax} and
consider a future-directed ($\Gamma > 0$) ingoing ($\sigma = -1$) or outgoing
($\sigma = +1$) geodesic, timelike or nonradial null (see examples in Figs.
\ref{timgeosm} and \ref{norad1}). Since $P^2 \leq 1$, $Q^2 \leq 1$, $E > 0$ and
$\phi \geq 0$, it is clear from (\ref{3.9}) -- (\ref{3.10}) that $\dril P s \geq
0$ and $\dril Q s \geq 0$ (equality only at $r = m$ and $r = \infty$) and they
cannot change sign, so such a geodesic will keep proceeding towards larger $P$
and larger $Q$ as long as $r \in (m, \infty)$.

Now consider a geodesic going off the same initial point E1 to the past ($\Gamma
< 0$) and towards decreasing ($\sigma = -1$) or increasing ($\sigma = +1$) $r$.
This time $\dril P s \leq 0$ and $\dril Q s \leq 0$, and they again cannot
change sign.

An ingoing ($\sigma = -1$) future-directed ($\Gamma > 0$) geodesic with the
initial point in sector I reaches the upper $r = m$ line in the right panel of
Fig. \ref{compact} with $Q \in (-1, +1)$ and $P = 1$. From (\ref{3.9}) --
(\ref{3.10}) and (\ref{2.7}) we have
\begin{equation}\label{3.11}
\dr V U = \frac {\left(1 - Q^2\right) E + \frac {1 - P^2} {\phi}\ \left(\Gamma +
\sqrt{\Gamma^2 - E \phi}\right)^2} {\left(1 - Q^2\right) E - \frac {1 - P^2}
{\phi}\ \left(\Gamma + \sqrt{\Gamma^2 - E \phi}\right)^2}\ .
\end{equation}
As shown in Appendix \ref{derVU},
\begin{equation}\label{3.12}
\lim_{r \to m} \frac {1 - P^2} {\phi} = 0.
\end{equation}
Hence
\begin{equation}\label{3.13}
\lim_{r \to m} \dr V U = +1.
\end{equation}
Thus, such a geodesic, timelike or null, radial or nonradial, in the $(U, V)$
coordinates is tangent to the horizon at the point of contact. The numerical
examples further on will confirm this.

\section{Examples of radial timelike geodesics}\label{numgeo}

\setcounter{equation}{0}

\begin{figure}
 \begin{center}
 ${}$ \\[-3cm]
 \hspace{-2cm}  \includegraphics[scale=0.6]{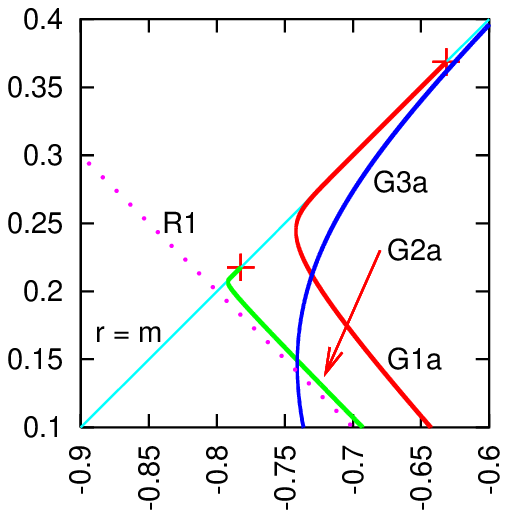}
 \end{center}
 ${}$ \\[-3.5cm]
 \hspace{-5mm} \includegraphics[scale=0.62]{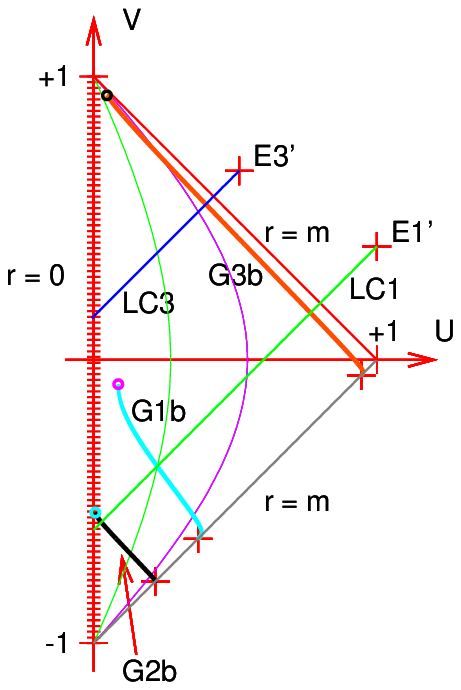}
 \includegraphics[scale=0.73]{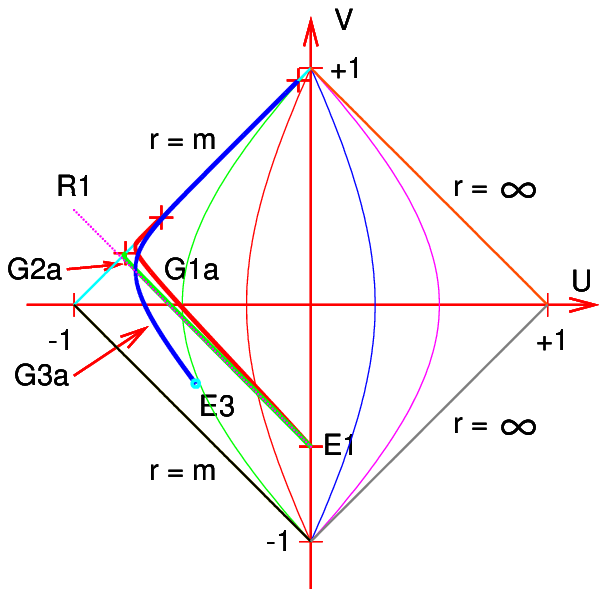}
 ${}$ \\[-1cm]
 \caption{
 \label{timgeosm}
 \footnotesize
{\bf Right panel:} Three future-directed radial timelike geodesics emitted at
points E1 and E3 in sector I of Fig. \ref{rnextmax}. The G1a has $\Gamma = 1.1$,
G2a has $\Gamma = 3.0$ (i.e. larger energy in the Newtonian limit), G3a is
`elliptic' and has $\Gamma = 0.5$. R1 is a radial ray emitted at E1; it hits the
singularity at $r = 0$ in a finite interval of the affine parameter. The crosses
at the upper ends of G1a, G2a and G3a mark the first points on them at which $r
< m$. {\bf Left panel:} The continuation of G1a, G2a and G3a into sector II of
Fig. \ref{rnextmax}. Their endpoints, marked with dots, are at the turning
points. LC1 is the radial generator of the past light cone of E1$'$ -- the copy
of E1 in sector I$'$ of Fig. \ref {rnextmax}, LC3 is the analogue of LC1 for
point E3$'$. See the text for more explanation. {\bf Upper inset:} An enlarged
image of the area around the upper endpoints of G1a and G2a. }
 \end{figure}

We now take the point E1 in sector I of Fig. \ref{rnextmax}, of coordinates $(U,
V) = (0, -0.6)$, as the initial point of two radial timelike ($E = 1$) ingoing
($\sigma = -1$) geodesics, future-directed ($\Gamma > 0$), with $\Gamma = 1.1$
in one example (the G1a in the right panel of Fig. \ref{timgeosm}) and $\Gamma =
3.0$ in another example (the G2a). We calculate the corresponding initial $(P,
Q)$ and $(p, q)$ from (\ref{2.7}) and (\ref{2.6}) and the initial $r_i =
1.8999999999292900$ via $\zeta$ from (\ref{2.5}) and (\ref{2.2}). We proceed
with step $\Delta s = 10^{-6}$ calculating $P(s)$ and $Q(s)$ from (\ref{3.9}) --
(\ref{3.10}), then $U(s)$ and $V(s)$ from (\ref{2.7}). As predicted in Sec.
\ref{geodesics} both these geodesics approach the horizon $r = m$ tangentially,
see Fig. \ref{timgeosm}.

In choosing the initial point E3 of the elliptic geodesic G3a one must ensure
that the initial $r = r_i$ is smaller than the $r_{{\rm TP}-}$ of (\ref{3.5})
(otherwise, $\Gamma^2 - E \phi < 0$ at the initial point, and the numerical
program will refuse to proceed). Consequently, $(t_i, r_i)$ are more convenient
initial data than $(U_i, V_i)$. Given $m = 0.95$, we thus choose
\begin{equation}\label{4.1}
(t_i, r_i) = (0.1, \ \ r_{{\rm TP}-} - 0.2).
\end{equation}
Then we calculate the initial $\zeta$ from (\ref{2.2}), the initial $(P, Q)$
from (\ref{2.6}) and the initial $(U, V)$ from (\ref{2.7}). From this point on,
we follow the scheme described in the preceding paragraph: we send the ingoing
radial timelike geodesic G3a towards the future, see the right panel of Fig.
\ref{timgeosm}.

The left panel of Fig. \ref{timgeosm} shows the continuation of the three
geodesics into sector II, these are G1b, G2b and G3b, respectively (see below
for a technical comment). Their upper endpoints are at their turning points. The
line marked LC1 is the radial generator of the past light cone of E1$'$, the
copy of point E1 in sector I$'$ of Fig. \ref{rnextmax}. The turning points of
G1b and G2b lie to the future of LC1. This means that if they were continued
beyond the turning points, they could not enter the causal past of E1$'$, so
they could not carry a message to the past of E1$'$, i.e. causality is not
broken in these two cases. The point E3$'$ is the copy of E3 in sector I$'$, the
line LC3 is the radial generator of the past light cone of E3$'$. One can see
that also for this geodesic, the turning point lies to the future of LC3, so
causality is not broken.

Here comes the technical comment: as stated earlier, G1a, G2a and G3a are
tangent to the horizon at the point of contact, so numerical approximation
errors cause that G1b, G2b and G3b cannot get off the horizon on the other side.
Therefore, the initial values of $V$ in sector II were corrected by $V \to V +
10^{-6}$. This correction is invisible at the scale of Fig. \ref{timgeosm}. See
Appendix \ref{corrections} for more comments and technical details.

\begin{figure}[h]
 \begin{center}
 \hspace{-4cm}
 \includegraphics[scale=0.75]{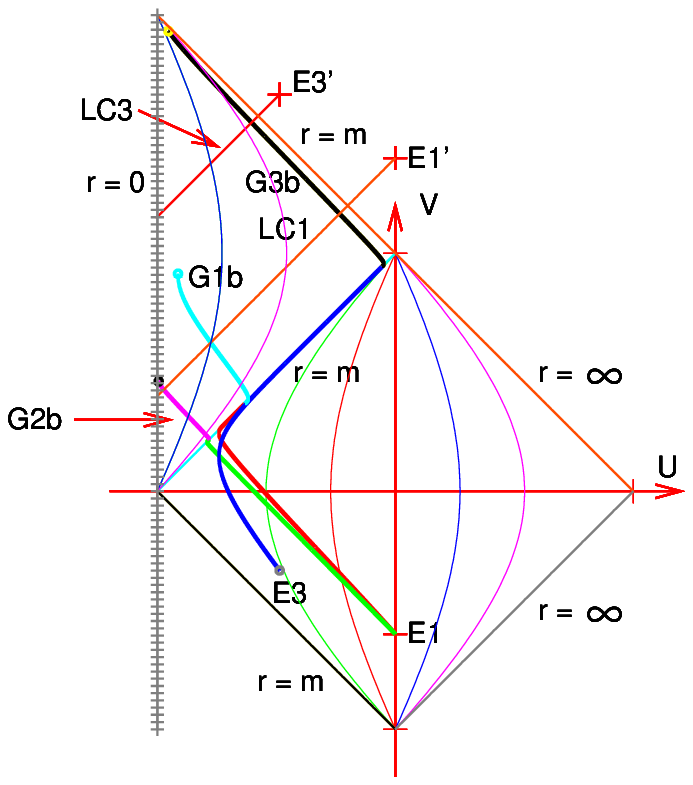}
 \caption{
 \label{wholetimgeo}
 \footnotesize
The two lower panels of Fig. \ref{timgeosm} put into their correct relative
positions. }
 \end{center}
 \end{figure}

Figure \ref{wholetimgeo} shows the two lower panels of Fig. \ref{timgeosm} in
their correct relative positions.

All three geodesics G1b, G2b and G3b have their turning points to the future of
the past light cones of the copies of their points of origin, E1 and E3. It is
thus clear that if they were continued beyond the turning points, they could not
enter the causal past of E1$'$ and E3$'$, respectively, so could not carry
messages from E1 and E3 to the past of E1$'$ and E3$'$. Consequently, the
identifications of the asymptotically flat regions do not lead to causality
breaches {\it in these numerical examples}. It remains an open problem to prove
by a formal mathematical reasoning that this is so for all geodesics.

\section{Nonradial timelike and null geodesics}\label{norad}

\setcounter{equation}{0}

The turning points of nonradial timelike or null geodesics are at the values of
$r$ that obey $\Gamma^2 - E \phi = 0$ in (\ref{3.2}) -- (\ref{3.3}), which is
equivalent to
\begin{equation}\label{5.1}
\Gamma^2 r^4 - \left(\varepsilon r^2 + {J_0}^2\right) \left(r^2 - 2mr +
m^2\right) = 0
\end{equation}
and also to
\begin{equation}\label{5.2}
\left(\Gamma^2 / E - 1\right) r^2 + 2mr - m^2 = 0.
\end{equation}
Equivalently, (\ref{5.2}) may be written as
\begin{equation}\label{5.3}
r = m \frac {1 - \sigma |\Gamma|/ \sqrt{E}} {1 - \Gamma^2 / E} \equiv \frac m {1
+ \sigma |\Gamma| / \sqrt{E}}, \qquad \sigma = \pm 1.
\end{equation}
This is still an equation to solve (because $E$ depends on $r$), but it it is
more useful than (\ref{5.1}) for a discussion. The fourth-degree equation
(\ref{5.1}) may have 0 to 4 real solutions for $r$. If it has any solutions,
then those with $\sigma = +1$ have $r < m$, and those with $\sigma = -1$ either
have no physical implications (because $r < 0$ or $r \to \pm \infty$ when
$|\Gamma| / \sqrt{E(r)} \geq 1$) or have $r > m$ (when $|\Gamma| / \sqrt{E(r)} <
1$). The latter are irrelevant for the problem of causality because they do not
enter the black hole region.

Equation (\ref{5.3}) has elementary solutions for null geodesics, for which
$\varepsilon = 0$ and so $E = {J_0}^2 / r^2$. For $\sigma = +1$, the solution is
\begin{equation}\label{5.4}
{\overset N r}{}_{\rm TP} = \frac 1 2\ \left|J_0/\Gamma\right| \left(-1 +
\sqrt{1 + 4 m \left|\Gamma / J_0\right|}\right) \equiv \frac {2m} {1 + \sqrt{1 +
4 m \left|\Gamma / J_0\right|}} < m
\end{equation}
(the other solution with $\sigma = +1$ in (\ref{5.3}) has minus in front of the
square root in (\ref{5.4}), so $r < 0$ and such a TP does not exist). The
solution (\ref{5.4}) exists for all values of $m$, $\Gamma$ and $J_0$, and
$\lim_{J_0 \to 0}{\overset N r}{}_{\rm TP} = 0$, which is consistent with our
earlier finding that a radial null geodesic can hit $r = 0$.

For $\sigma = -1$, two extra solutions of (\ref{5.3}) (i.e. two additional TPs)
exist when
\begin{equation}\label{5.5}
4 m |\Gamma / J_0| < 1.
\end{equation}
For these, ${\overset N r}{}_{\rm TP} > m$ as announced, but they create an
interesting situation. They are
\begin{equation}\label{5.6}
\left[\begin{array}{l} {\overset N r}{}_{\rm TP4} \\
         \\
                       {\overset N r}{}_{\rm TP5} \end{array}\right] =
\left[\begin{array}{l} {\displaystyle{\frac {2m} {1 - \sqrt{1
                                   - 4m \left|\Gamma/J_0\right|}}}} \\
        \\
                       {\displaystyle{\frac {2m} {1 + \sqrt{1
                                  - 4m \left|\Gamma/J_0\right|}}}} \\
\end{array}\right].
\end{equation}
Then, the set of of nonradial null geodesics splits into two families. In one,
the geodesics (light rays) move between ${\overset N r}{}_{\rm TP}$ and
${\overset N r}{}_{\rm TP5}$, in the other they move between ${\overset N
r}{}_{\rm TP4}$ and infinity. The geodesics of the second family never leave
sector I, and so are irrelevant for the problem of causality. But the geodesics
of the first family cross the horizon from sector I into sector II, and then
continue to sector I$'$. These are relevant, and they will be mentioned again in
the next section.

When $4m |\Gamma / J_0| = 1$, we have ${\overset N r}{}_{\rm TP5} = {\overset N
r}{}_{\rm TP4} \df {\overset N r}{}_{\rm TP}$. Then the geodesics of the two
families mentioned above approach the TP from opposite sides and bounce -- one
towards infinity, the other towards the horizon. This situation is qualitatively
not much different from that with  $4m |\Gamma / J_0| < 1$.

\section{Numerical examples of nonradial geodesics}\label{numnorad}

\setcounter{equation}{0}

\begin{figure}
 \begin{center}
 ${}$ \\[-5cm]
 \hspace{-3cm} \includegraphics[scale=0.6]{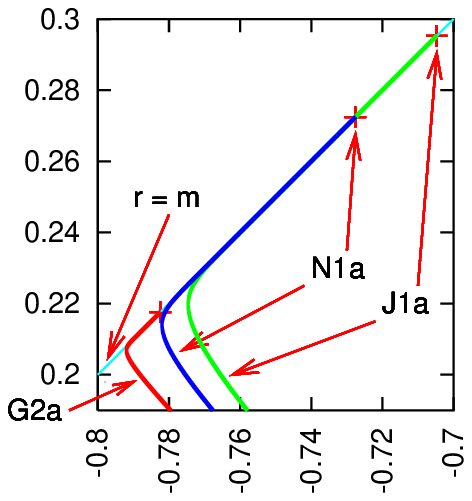}
 \end{center}
 ${}$ \\[-3cm]
 \hspace{-4cm}\includegraphics[scale=0.6]{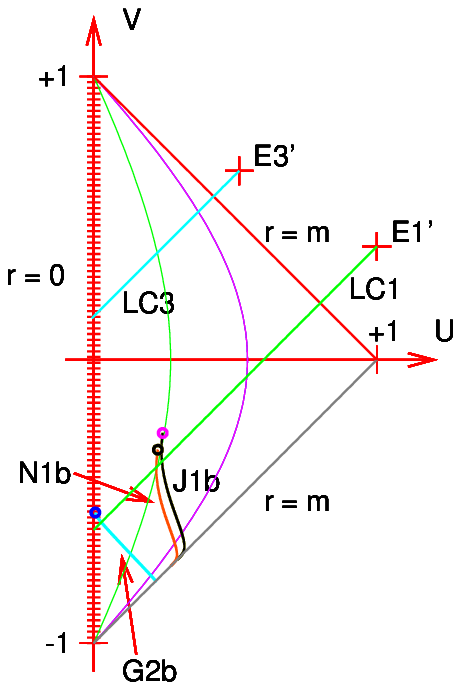}
 \includegraphics[scale=0.7]{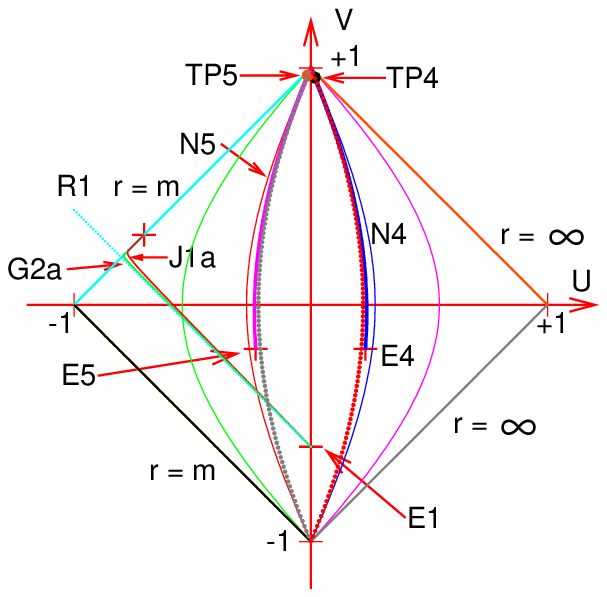}
 \caption{
 \label{norad1}
 \footnotesize
{\bf Right panel:} The radial geodesic G2a and the image of the nonradial
timelike geodesic J1a going off point E1 in sector I of Fig. \ref{rnextmax}. At
this scale they nearly coincide. The image of the nonradial null geodesic N1a is
squeezed between them and invisible. The cross nearly on $r = m$ marks the
endpoint of J1a (the first point at which $r < m$). {\bf The upper inset:} A
closeup view on the neighbourhood of the points where the geodesic G2a and the
images of N1a and J1a cross the horizon. The crosses that seemingly lie on $r =
m$ mark the first points on the geodesics at which $r < m$. {\bf Left panel:}
The continuation of G2a and of the images of N1a and J1a into sector II of Fig.
\ref{rnextmax}. The turning points of the nonradial geodesics are above the past
light cone of E1$'$. See the text for explanations concerning the other curves
and points in the right panel. }
 \end{figure}

We now numerically calculate a nonradial timelike geodesic J1a and a nonradial
null geodesic N1a that cross the horizon outside in, with the initial point in
sector I. We take
\begin{equation}\label{6.1}
\left(m, \Gamma, J_0\right) = (0.95, \quad 3.0, \quad 2.6);
\end{equation}
the first parameter as in this whole paper, the second as for the radial
geodesic G2a, the third one chosen at random, and we take the same E1 as before
for the initial point with the coordinates $(U, V) = (0.0, -0.6)$. With $J_0
\neq 0$ the geodesics do not stay in the initial $(U, V)$ plane, but go around
the $V$ axis. For comparing them with the radial one, we rotate each of their
points around the $r = 0$ (i.e., $U = -1$) axis into the $(U, V)$ plane. (This
happens simply by placing the $(U, V)$ coordinates of a point P in the $(U, V)$
plane and ignoring the fact that P has a nonzero $\varphi$. An illustration is
given at the end of this section.) The comparison of those projections with the
radial G2a is shown in Fig. \ref{norad1}. Both projections are close to G2a
throughout sector I. While crossing the horizon, the projections of J1a and N1a
are offset further than G2a, but reach the turning point also above the past
light cone of E$'$. The projection of N1a stays between that of J1a and G2a and,
for better transparency, is not shown in sector I.

For illustration, the lower right panel of Fig. \ref{norad1} shows also two
nonradial null geodesics N4 (initially ingoing) and N5 (initially outgoing),
whose parameters obey (\ref{5.5}):
\begin{equation}\label{6.2}
\left(m, \Gamma, J_0\right) = (0.95, \quad 3.0, \quad 11.41).
\end{equation}
They are examples of the nonradial null geodesics with large $|J_0|$, discussed
in connection with (\ref{5.5}) and (\ref{5.6}). The $r$ coordinates of their
extra turning points are
\begin{equation}\label{6.3}
\left({\overset N r}{}_{\rm TP5}, {\overset N r}{}_{\rm TP4}\right) =
(1.8453688474864800, \quad 1.9579644858468537).
\end{equation}
The dotted arcs are at $r = {\overset N r}{}_{\rm TP5}$ (the left one) and $r
={\overset N r}{}_{\rm TP4}$ (the right one). The initial point E4 of N4 has
coordinates
\begin{equation}\label{6.4}
(t, r) = (-0.2, \quad {\overset N r}{}_{\rm TP4} + 0.005).
\end{equation}
The N4 goes towards smaller $r$ until it reaches the turning point TP4. Then it
becomes outgoing and recedes to infinity. The $(U, V)$ coordinates of TP4 are
\begin{equation}\label{6.5}
(U, V)_{\rm TP4} = (0.016897625562350471, \quad 0.95928250934355219).
\end{equation}
The other geodesic, N5, goes off point E5, with coordinates
\begin{equation}\label{6.6}
(t, r) = (-0.2, \quad {\overset N r}{}_{\rm TP5} - 0.005).
\end{equation}
and is initially outgoing. It reaches the turning point TP5 at $r = {\overset N
r}{}_{\rm TP5}$, where it becomes ingoing. The $(U, V)$ coordinates of TP5 are
\begin{equation}\label{6.7}
(U, V)_{\rm TP5} = (-0.012965471226367431, \quad 0.96888417351466494).
\end{equation}
Unfortunately, N5 crosses the horizon so near to $(U, V) = (0, 1)$ that its
continuation into sector II could not be calculated even at double precision in
Fortran 90.

Another null geodesic, with the same parameters as N5 and the same initial point
E5 but ingoing from the start, coincided with N5 at the scale of Fig.
\ref{norad1} and had the point of contact with the horizon also very near to
$(U, V) = (0, 1)$.

\begin{figure}[h]
 \begin{center}
 ${}$ \\[5mm]
 \hspace{-2cm}
 \includegraphics{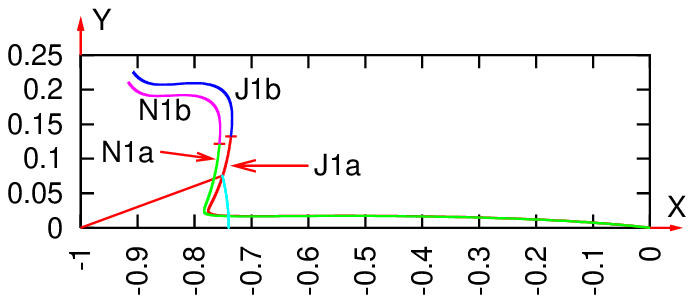}
 \caption{
 \label{proj}
 \footnotesize
The view on nonradial geodesics from atop the $r = 0$ axis in Fig.
\ref{wholetimgeo}. See the text for explanation.}
 \end{center}
 \end{figure}

As a curiosity, Fig. \ref{proj} shows the view on geodesics J1a, J1b, N1a and
N1b from atop the $r = 0$ axis in Fig. \ref{wholetimgeo}. The radial coordinate
here is $(U + 1)$ with $U$ as in Fig. \ref{wholetimgeo}, the origin is at $U =
-1$, the $V$ axis goes perpendicularly to the figure plane towards the viewer,
and
\begin{equation}\label{6.8}
X = (U + 1) \cos \varphi - 1, \qquad Y = (U + 1) \sin \varphi.
\end{equation}
The initial point (the right end of J1a and N1a) is at $(U, \varphi) = (0, 0)$.
The consecutive values of $\varphi$ were calculated using (\ref{3.4}). The short
horizontal bars are where the geodesics cross the horizon. The inclined straight
segment and the arc at its end show how the projections of the points of the
geodesics to the plane of Fig. \ref{norad1} were constructed. The projections of
J1a and N1a nearly coincide between $X = 0$ and $X = -0.78$.

\begin{figure}[h]
 \begin{center}
 \hspace{-2cm}
 \includegraphics[scale=0.8]{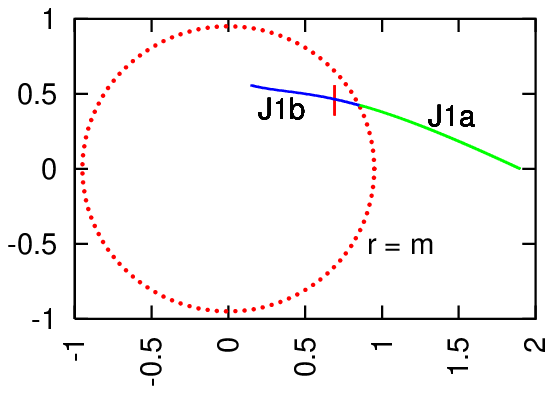}
 \caption{
 \label{rproj}
 \footnotesize
The view on the geodesics J1a and J1b analogous to Fig. \ref{proj}, but in the
$(x, y) = (r \cos \varphi, r \sin \varphi)$ coordinates. See the text for a
comment.}
 \end{center}
 \end{figure}

Finally, Fig. \ref{rproj} shows the view on J1a and J1b analogous to that in
Fig. \ref{proj}, but in the $(x, y) = (r \cos \varphi, r \sin \varphi)$
coordinates. In this projection, N1a and N1b nearly coincide with J1a and J1b,
so they are not shown. As can be seen, in these coordinates the geodesic is not
tangent to the horizon. This agrees with (\ref{3.2}), which shows that $\dril r
s = \pm \Gamma \neq 0$ at $r = m$. Drawing Fig. \ref{rproj} required a trick
that is explained in Appendix \ref{trick}.

\section{Summary and conclusions}\label{conclu}

\setcounter{equation}{0}

The aim of this paper was to verify whether an observer in the maximally
extended extreme RN spacetime with asymptotically flat regions (AFRs) identified
can send messages to its own past by means of timelike or null geodesics (in
other words, whether the identifications of the AFRs lead to acausality). By
this opportunity, images of geodesics in this extension were derived and
discussed.

In Sec. \ref{RNintro}, the maximal extension of the extreme RN metric was
re-derived in more detail than in standard textbook presentations. In
particular, in the coordinates of the extension chosen here, the shapes of the
singular set $r = 0$ and of the lines of constant $t$ and $r$ were explicitly
calculated.

In Sec. \ref{geodesics}, the geodesic equations in the extreme RN metric were
discussed. It was shown that radial null geodesics (NGs) can hit the singularity
at $r = 0$, and those that do stop there. Timelike and nonradial NGs that cross
the horizon $r = m$ are tangent to it in the $(U, V)$ coordinates of the maximal
extension. The $r$-coordinate of the turning point (TP) of a radial timelike
geodesic (TG) is given by the simple Eq. (\ref{3.5}).

In Sec. \ref{numgeo}, three examples of radial TGs were numerically calculated,
with different values of the energy constant $\Gamma$. As predicted, they cross
the horizon tangentially. Their TPs lie to the future of the past light cones
(PLCs) of the copies of their initial points (the copies are created by
identifying the asymptotically flat regions). Therefore, they cannot carry
messages to the past of their emitters, i.e. the identifications of the AFRs do
not lead to causality breaches.

In Sec. \ref{norad}, general properties of nonradial TGs and NGs were discussed.
For NGs, the $r$-coordinates of their TPs are given by explicit exact formulae,
and so they were discussed in more detail. One TP exists for every nonradial NG
and lies inside the horizon. With sufficiently large $|J_0|$ (where $J_0$ is the
angular momentum constant), two extra TPs exist outside the horizon. In this
case, the nonradial NGs move either between the outermost TP and the infinity,
or between the other two TPs. In the second case, they cross the horizon and go
into the next AFR.

In Sec. \ref{numnorad}, numerical examples of the nonradial geodesics that cross
the horizon were presented, one timelike (named J1) and one null (named N1). For
them, too, the TPs lie to the future of the PLCs of the copies of their emission
points, so causality is not broken. Two numerical examples of nonradial NGs with
large $|J_0|$ that illustrate the calculations of Sec. \ref{norad} were also
presented.  In addition, the projections of J1 and N1 on surfaces of constant
$V$, one in the $(U, \varphi)$ coordinates and one in the $(r, \varphi)$
coordinates, were shown in illustrations.

In contrast to the RN metric with $e^2 < m^2$, in the $e^2 = m^2$ case all
numerical examples show the same:

\bigskip

\parbox{5in}{
\ \ \ \ Let E be the initial point of a geodesic in the asymptotically flat
region I of the maximally extended extreme RN spacetime. Let E$'$ be the copy of
E in the first future copy of I. A timelike or nonradial null geodesic emitted
at E will have its turning point outside the past light cone of E$'$. Thus, a
message sent by this geodesic will not reach the causal past of E$'$. This means
that the identification of E$'$ with E does not cause acausality. }

\bigskip

However, it remains an open problem to prove the same by a formal mathematical
reasoning. Nonexistence (here: of geodesics breaking causality) cannot be proved
by examples alone.

\appendix

\section{The singularity of (\ref{2.1}) at $r = m$ is spurious}\label{spusing}

\setcounter{equation}{0}

The orthonormal tetrad of differential forms connected with the metric
(\ref{2.1}) is
\begin{equation}\label{a.1}
e^0 = (1 - m / r) {\rm d} t, \quad e^1 = (1 - m / r)^{-1} {\rm d} r, \quad e^2 =
r {\rm d} \vartheta, \quad e^3 = r \sin \vartheta {\rm d} \varphi.
\end{equation}
The independent nonzero tetrad components of the Riemann tensor in this tetrad
are
\begin{eqnarray}\label{a.2}
R_{0101} &=& 2m / r^3 - 3m^2 / r^4, \qquad R_{2323} = - 2m / r^3 + m^2 / r^4
\nonumber \\
R_{0202} &=& R_{0303} = - R_{1212} = - R_{1313} = - \frac m {r^3} \left(1 -
\frac m r\right).
\end{eqnarray}
They are all regular at $r = m$, so there is no curvature singularity at this
$r$. $\square$

\section{Solving (\ref{2.2}) numerically for $r$ given
$\zeta(r)$}\label{major}

\setcounter{equation}{0}

While numerically integrating the geodesic equations, we are confronted with the
problem of determining $r$ from (\ref{2.2}) for a given $\zeta$. This is done by
the bisection method, separately in the $r \in (0, m)$ and in the $r \in (m,
\infty)$ domain. In $(0, m)$, the initial bounds for $r$ are obviously $0 < r <
m$. In $(m, \infty)$, the lower bound is $r = m$, but the upper bound is not
self-evident. This is how it can be determined.

\begin{figure}[h]
 \hspace{-5mm} \includegraphics[scale=0.6]{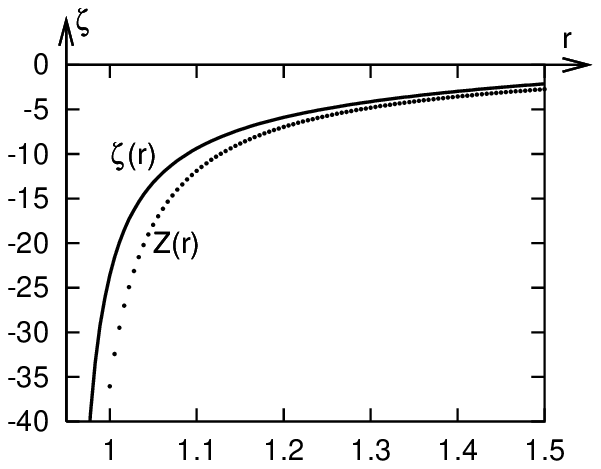}
 \hspace{5mm} \includegraphics[scale=0.6]{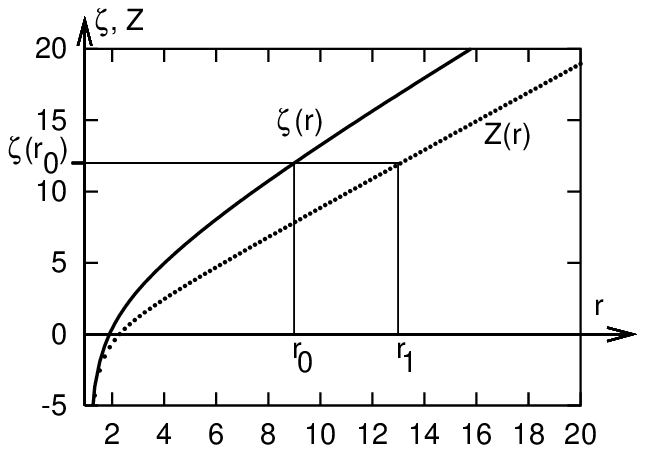}
 \caption{
 \label{majorfig}
 \footnotesize
Comparison of the functions $\zeta(r)$ and $Z(r)$ for small $r > m$ (left panel)
and for large $r$ (right panel). The right panel shows how to determine the
initial $r_1$ in finding $r_0$ for a given $\zeta(r_0)$. }
 \end{figure}

Since the function $\zeta(r)$ changes monotonically in the whole $(- \infty, +
\infty)$ range, we have to find a function $Z(r)$ that has the same range, $Z(r)
< \zeta(r)$ for all $r \in (m, \infty)$ and $Z(r_1) = Z_1$ is easy to solve for
$r_1$ given $Z_1$. For every $x > 0$ we have $2 \ln x > - 1 / x$ (the proof is
left as an exercise for the reader). Hence, for $r > m$ we have in (\ref{2.2})
\begin{equation}\label{b.1}
2m \ln |r/m - 1| > - m / (r / m - 1) \equiv - m^2 / (r - m) \quad {\rm for\ all}
\ \ r > m.
\end{equation}
Consequently,
\begin{equation}\label{b.2}
\zeta(r) > Z(r) \df r - m - 2 m^2 / (r - m) \quad {\rm for\ all} \ \ r > m.
\end{equation}
Thus, given $\zeta_0 \df \zeta(r_0)$, where $r_0$ is to be found, we solve
$Z(r_1) = \zeta_0$ for $r_1$ and find
\begin{equation}\label{b.3}
r_1 = m + \frac 1 2\ \left(\zeta_0 + \sqrt{{\zeta_0}^2 + 8 m^2}\right)
\end{equation}
(the other solution, with ``$-$'' in front of $\sqrt{\ }$, would have $r_1 <
m$). By construction, $r_0 < r_1$. This calculation is illustrated in Fig.
\ref{majorfig}.

\section{Christoffel symbols in the $(t, \zeta)$ coordinates}\label{Christ}

\setcounter{equation}{0}

The symbol $S_1$ stands for
\begin{equation}\label{c.1}
S_1 = 1 - m / r,
\end{equation}
so $S_1 = 0$ at $r = m$. The formulae below show that there is no singularity in
the Christoffel symbols there (in the coordinates of (\ref{2.1}), some
Christoffel symbols contain the factor $S^{-1}$). Only the independent nonzero
components of $\tChr {\alpha\ } {\beta} {\gamma}$ are shown; $(x^0, x^1, x^2,
x^3) = (t, \zeta, \vartheta, \varphi)$.

\begin{eqnarray}
&& \Chr {0\ } {0} {1} = \Chr {1\ } {0} {0} = \Chr {1\ } {1} {1} = mr^{-2}{S_1},
\\
&& \Chr {1\ } {2} {2} = \Chr {1\ } {3} {3} / {\sin}^{2}(\vartheta) = - r, \\
&& \Chr {2\ } {1} {2} = \Chr {3\ } {1} {3} = r^{-1}{S_1}^{2}, \\
&& \Chr {2\ } {3} {3} = - \cos (\vartheta)\sin (\vartheta), \qquad \Chr {3\ }
{2} {3} = \cos(\vartheta) / \sin(\vartheta).
\end{eqnarray}

\section{The limit of $\dril V U$ at $r \to m$ with $\sigma = -1$ and $\Gamma >
0$}\label{derVU}

\setcounter{equation}{0}

Since $P = 1$ and $\phi = 0$ at $r = m$, we have in (\ref{3.11})
\begin{equation}\label{d.1}
\lim_{r \to m} \frac {1 - P^2} {\phi} = \lim_{r \to m} \left[\frac {-2 P \dril P
s} {2 (1 - m / r) (m / r^2) \dril r s}\right].
\end{equation}
Using (\ref{3.2}) with $\dril r s < 0$ and (\ref{3.9}) with $\sigma = -1$, we
obtain from this
\begin{equation}\label{d.2}
\lim_{r \to m} \frac {1 - P^2} {\phi} = \lim_{r \to m} \left[\frac {P
\left(\Gamma + \sqrt{\Gamma^2 - E \phi}\right)} {(1 - m / r) (m / r^2)
\sqrt{\Gamma^2 - E \phi}} \times \frac {1 - P^2} {\phi}\right].
\end{equation}
Hence,
\begin{equation}\label{d.3}
\lim_{r \to m} \left\{\frac {1 - P^2} {\phi} \left[1 - \frac {r^2 P} {m (1 - m /
r)}\ \left(\frac {\Gamma} {\sqrt{\Gamma^2 - E \phi}} + 1\right)\right]\right\} =
0.
\end{equation}
But the second term in square brackets clearly tends to $(\mp \infty)$ as $r \to
m^{\pm}$. So, (\ref{d.3}) can hold only when
\begin{equation}\label{d.4}
\lim_{r \to m} \frac {1 - P^2} {\phi} = 0. \qquad \square
\end{equation}

\section{Crossing the horizon with numerical integration of the geodesic
equations}\label{corrections}

\setcounter{equation}{0}

\begin{figure}[h]
 ${}$ \\[-4cm]
 \hspace{-4cm} \includegraphics[scale=0.65]{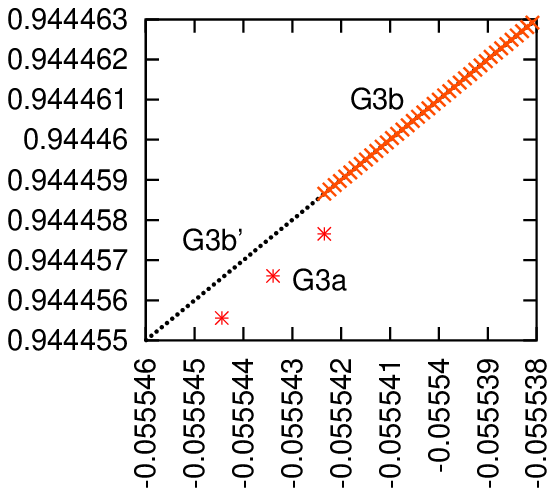}
 \hspace{-5mm} \includegraphics[scale=0.65]{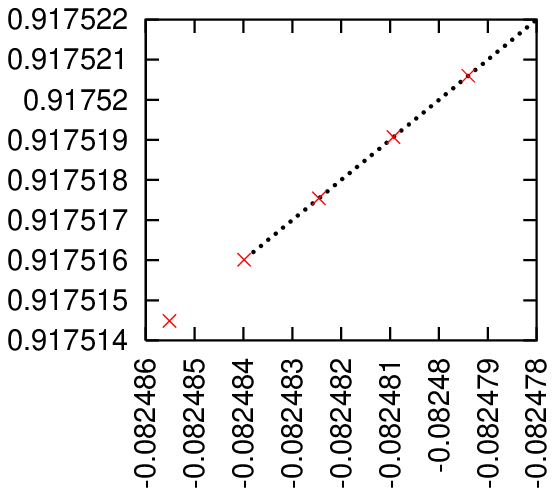}
 \caption{
 \label{G3corr}
 \footnotesize
{\bf Left panel:} The neighbourhood of the upper endpoint of G3a. The
artificially introduced jump $\Delta V = 10^{-6}$ in the $V$ coordinate between
the endpoint of G3a and the initial point of G3b is visible. The G3b$'$ is the
result of backtracking G3b from its upper end in Fig. \ref{timgeosm} to the
intersection with $r = m$. {\bf Right panel:} The neighbourhood of the lower
endpoint of G3b$'$, where it crosses $r = m$. Here, the $V$ coordinates of
points on G3b$'$ (drawn with dots) coincide with those on G3a (drawn with
crosses) to better than $10^{-6}$.}
 \end{figure}

As already stated in Sec. \ref{numgeo}, the G1a, G2a and G3a curves in Fig.
\ref{timgeosm} approach the horizon tangentially. When their points of contact
with $r = m$ were used as the initial points of their continuations into sector
II, numerical imprecisions caused that the continuations kept going along $r =
m$. The initial coordinates of the continuations had to be hand-corrected to
manageable values. Here we demonstrate the consequences of this correction for
the G3a,b geodesic, see Fig. \ref{G3corr}.

Recall: the G3a was numerically integrated from point E3 to the first point
(call it P) at which $r$ became smaller than $m$. The $V_P$ coordinate of P was
increased `by hand' to $V'_P = V_P + 10^{-6}$. Then, the $(U_P, V'_P)$
coordinates of P in sector I were transformed to $(\widetilde{U}, \widetilde{V})
= (1 + U_P, V'_P - 1)$ -- the coordinates of the same point in sector II. Using
$(\widetilde{U}, \widetilde{V})$ as the initial data, the geodesic equations
were integrated up to the turning point, at which $r = r_{{\rm TP}+}$ given by
(\ref{3.5}); this second segment of G3a is denoted G3b. The $(U, V)$ coordinates
of the turning point were
\begin{equation}\label{e.1}
(U_f, V_f) = (0.043718751055178640, \quad 0.93765063538654081).
\end{equation}
To verify the precision of the code, a past-directed radial timelike geodesic
G3b$'$ was sent from $(U_f, V_f)$ backward with the same $\Gamma$ (backward
means towards increasing $r$, i.e. with $\sigma = +1$ in (\ref{3.9}) --
(\ref{3.10})). The endpoint of G3b$'$ was where $r$ became larger than $m$.
Figure \ref{G3corr} shows the relations between G3a, G3b and G3b$'$ in the
neighbourhood of point P. The left panel shows that at P G3b$'$ coincides with
G3b to better than $10^{-6}$; it also shows the jump between G3a and G3b. The
right panel shows that at its lower end G3b$'$ coincides with G3a to better than
$10^{-6}$.

\section{Drawing Fig. \ref{rproj}}\label{trick}

\setcounter{equation}{0}

The hand-correction described at the beginning of Appendix \ref{corrections}
caused a visible jump in $r$. The geodesic J1b had its initial point in Fig.
\ref{rproj} at the short vertical bar. To close the gap, another geodesic J1b$'$
was issued from the future endpoint of J1b backward, and allowed to cross the $r
= m$ circle. This is the arc marked J1b in Fig. \ref{rproj}; it coincides with
the proper J1b between the bar and the left endpoint.

In the first calculation it was assumed that the first value of $\varphi$ on J1b
is the same as the last value on J1a, which caused another (small) discontinuity
between J1a and J1b. Consequently, to make J1a and J1b$'$ meet with the
precision of $\Delta y = 10^{-5}$ shown in the figure, the correction $\Delta
\varphi = 0.011826$ had to be applied to the value of $\varphi$ at the left end
of J1b$'$. It was determined by trial and error. The value of $\varphi$ does not
appear in (\ref{3.1}) -- (\ref{3.3}), so such manipulations with it did not
require any change in the numerical algorithm of calculating the geodesics.

{\bf Acknowledgement.} For some calculations, the computer algebra system
Ortocartan \cite{Kras2001,KrPe2000} was used.

\end{document}